\documentstyle[aps,prb,twocolumn,epsfig]{revtex}
\begin{document}

\twocolumn[\hsize\textwidth\columnwidth\hsize\csname
@twocolumnfalse\endcsname

\title{Temperature-scaling behavior of the Hall conductivity for 
	  Hg-based superconducting thin films} 
\author{Wan-Seon Kim, W. N. Kang, S. J. Oh, Mun-Seog Kim, Yujoung Bai, and Sung-Ik Lee}
\address{National Creative Research Initiative Center for Superconductivity and 
Department of Physics, Pohang University of Science and Technology, 
Pohang 790-784, Korea}
\author{Chang Ho Choi and H.-C. Ri}
\address{Material Science Team, Joint Research Division, Korea Basic Science
Institute, Taejon 305-333, Korea}
\maketitle
\draft
\begin{abstract}                                   
The Hall conductivities of HgBa$_{2}$CaCu$_{2}$O$_{6+\delta}$ 
and HgBa$_{2}$Ca$_{2}$Cu$_{3}$O$_{8+\delta}$
thin films are investigated for a magnetic field  
parallel to the $\it c$ axis. 
The mixed-state Hall conductivity for these compounds is well described by 
$\sigma_{xy}=C_{1}/H+C_{2}+C_{3}H$.
The prefactor $C_{1}$ shows a temperature dependence 
of the form $C_{1}\propto (1-t)^{n}$ near $T_{c}$, 
where $t=T/T_{c}$ is the reduced temperature.
Contrary to the previous results, 
$C_{2}$ also follows a temperature-scaling behavior similar to that 
of the coefficient $C_{1}$.
The observed value of $n$, $1.8 \sim 2.3$, is comparable to the previously observed 
values for YBa$_{2}$Cu$_{3}$O$_{7-\delta}$ and La$_{2-x}$Sr$_x$CuO$_4$. 
\end{abstract}                                   


\vskip 0.5pc]

\section{INTRODUCTION}

The Hall effect in the mixed state of high-temperature superconductors (HTS)
is one of the most interesting and controversial problems related to vortex
dynamics. Many experiments have shown that the Hall anomaly occurs not only
in HTS, such as YBa$_{2}$Cu$_{3}$O$_{7-\delta}$ (YBCO)\cite
{Hagen1990,Chien1991,Rice1992}, Bi$_{2}$Sr$_{2}$CaCu$_{2}$O$_{8}$ (Bi-2212)
\cite{Samoilov1993}, and Tl$_{2}$Ba$_{2}$CaCu$_{2}$O$_{8}$ (Tl-2212)\cite
{Hagen1991,Samoilov1994}, but also in conventional superconductors, for
example, in the thin-film and the single-crystalline forms of Nb\cite
{Hagen1990,Noto1976,Usui1968}, V\cite{Usui1968}, and In-Pb alloys\cite
{Weijsenfeld1968}.

So far, two kinds of sign reversals have been observed. The first is a
simple, single sign reversal as observed in YBCO\cite{Rice1992,Samoilov1995}
and La$_{2-x}$Sr$_x$CuO$_4$(LSCO)\cite{Matsuda1995}, and the second is a
double sign reversal from positive to negative and then to positive again as
the temperature decreases. The distinction between the two is that the
single sign reversal is observed for the case of relatively low anisotropy
while the double sign reversal is observed in the case of relatively high
anisotropy, such as Bi-\cite{Samoilov1993}, Tl-\cite{Hagen1991,Samoilov1995}%
, and Hg-based compounds\cite{WNKang1997,WNKang1999}. The anisotropy ratio
is known to be on the order of $10^{4}$ for Tl-2212, $10^{2}\sim 10^{3}$ for
LSCO, and $10\sim 10^{2}$ for YBCO, and the ratio for Hg-based
superconductors\cite{YCKim1995} is between the values for Tl-2212 and YBCO.

Recently, even a third sign reversal was observed in the low-temperature
region for heavy-ion-irradiated Hg-based compounds\cite{WNKang1999c}. This
observation is quite meaningful because this multiple sign reversal was
predicted by Kopnin and depended on the behaviors of the density of states
and of the gap of the superconductor\cite{Kopnin1996}. This multiple sign
reversal is possible if there are localized or almost localized energy
states in the superconducting state.

Just after the detection of the Hall effect in Nb crystals\cite{Reed1965},
Bardeen and Stephen\cite{Bardeen1965} derived the flux-flow resistivity and
the Hall resistivity due to vortex motion. However, in their theory, the
sign of the Hall resistivity is always positive. Quite a few other theories,
based on the flux-backflow\cite{Hagen1990}, two-band\cite{Hirsch1991}, or
induced-pinning\cite{Wang1994} phenomena, have also been developed to
explain the Hall effect in mixed states. However, the origin of the Hall
anomaly is still not well understood.

In this paper, we report the magnetic-field dependence of the Hall
conductivity in the mixed state of HgBa$_{2}$CaCu$_{2}$O$_{6+\delta}$
(Hg-1212) and HgBa$_{2}$Ca$_{2}$Cu$_{3}$O$_{8+\delta}$ (Hg-1223) thin films.
As expected, a double sign reversal is observed in these highly anisotropic
superconductors. The measured Hall conductivities in the mixed states are
better fitted by the form $\sigma_{xy}=C_{1}/H+C_{2}+C_{3}H$, which is
different from the case of the less anisotropic YBCO superconductor\cite
{Ginsberg1995} where $C_{2}$ is negligible. In this Hg-based superconductor, 
$C_{1}$ and $C_{2}$ depend strongly on the temperature, but that is not the
case for $C_{3}$. $C_{1}$ scales as $\sim (1-t)^{n}$, which is partially
understood from the temperature dependence of the gap and the coupling
constant\cite{Kopnin1996}, but this understanding is not rigorous. Here, we
claim that $C_{2}$, which appears only for highly anisotropic materials,
scales as $C_{2}\sim (1-t)^{n^{\prime}}$. The critical exponent $n^{\prime}$
is $2.0\pm 0.2$ for Hg-1223 and $3.2\pm 0.1$ for Hg-1212. We observe, for
the first time to the best of our knowledge, this scaling behavior of $C_{2}$
for Hg-based superconductors. A similar behavior was previously observed in
LSCO\cite{Matsuda1995}, but was not analyzed. $C_{3}$ for these Hg-based
thin films weakly depends on the temperature, which is a different behavior
than those observed for YBCO and LSCO.

\section{THEORETICAL BACKGROUD}

Kopnin {\it et al.}\cite{Kopnin1993} and Dorsey\cite{Dorsey1992b} obtained
the Hall conductivity by using the time-dependent Ginzburg-Landau (TDGL)
theory in which the relaxation time of the order parameter was taken to be
complex. Kopnin {\it et al.}\cite{Kopnin1993} claimed that the negative Hall
effect was very much related to the energy derivative of the density of
states. According to their theory, the Hall conductivity can be expressed by
two contributions. The first contribution due to the vortex motion is
proportional to $1/H$ and is dominant in the low-field region. The second
contribution, which originates from quasiparticles, is proportional to $H$.

An analysis of the Hall conductivity, based on the TDGL theory, for the YBCO
single crystal was reported by Ginsberg and Manson\cite{Ginsberg1995}. In
their paper, the Hall conductivity $\sigma_{xy}(H)$ was well explained by
the sum of $H$- and $1/H$-dependent parts. However, the behavior of $%
\sigma_{xy}(H)$ varies on a case-by-case basis. For example, the $H$%
-dependent part for YBCO is replaced by a field-independent part for Tl-2212
\cite{Samoilov1994}. In the case of LSCO\cite{Matsuda1995}, the Hall
conductivity is expressed as the sum of three terms : a $1/H$-dependent
term, an $H$-dependent term, and an $H$-independent term. The temperature
dependence of the coefficient of each component in the Hall conductivity was
also investigated\cite{Matsuda1995,Ginsberg1995}. The coefficient of the $1/H
$ term varies as $(1-t)^{n}$, where $t=T/T_{c}$ is the reduced temperature. $%
n$ is observed to be 2 for YBCO and $2\sim 3$ for LSCO.

Recently, Kopnin {\it et al.}\cite{Kopnin1996,Kopnin1995} calculated the
Hall conductivity based on the kinetic equations and the TDGL theory. Their
approach included an additional force due to the kinetic effects of changing
the quasiparticle densities in the normal core and in the superconducting
state. The total Hall conductivity is given by 
\begin{equation}
\sigma_{xy}(H)=\sigma_{H}^{(L)}+\sigma_{H}^{(D)}+\sigma_{H}^{(A)} ,
\label{sigma1}
\end{equation}
where $\sigma_{H}^{(L)}$ comes from localized excitations in the vortex
cores, $\sigma_{H}^{(D)}$ is from delocalized quasiparticles above the gap,
and $\sigma_{H}^{(A)}$ is from the additional force due to the kinetic
effects of charge imbalance relaxation. In the vicinity of $T_{c}$, this
term can be expressed as 
\begin{equation}
\sigma_{H}^{(A)} \sim \frac{1}{H \lambda} \left( \frac{d\nu}{d\zeta} \right)
\Delta^{2} ,  \label{sigmaA}
\end{equation}
where ${d\nu}/{d\zeta}$ is the energy derivative of the density of states at
the Fermi surface, $\lambda$ is the coupling constant, and $\Delta$ is the
superconducting energy gap. $\sigma^{(L)}_{H}$ and $\sigma^{(A)}_{H}$ depend
on $1/H$, while $\sigma^{(D)}_{H}$ is proportional to $H$. One important
thing we have to notice is that first two terms on the right-hand side of
Eq. (\ref{sigma1}) are always positive. For the dirty case, $\sigma^{(L)}_{H}
$ is very small; hence, it can be neglected near $T_{c}$. A sign reversal
can occur when $\sigma_{H}^{(A)}$ dominates over $\sigma_{H}^{(D)}$.

\section{EXPERIMENTALS}

High-quality Hg-1212 and Hg-1223 thin films were grown by using the pulsed
laser deposition and post annealing method. The details are reported
elsewhere\cite{WNKang1998,WNKang1999b}. The onset-transition temperatures, $%
T_c$, are 127 K for Hg-1212 and 132 K for Hg-1223. The sizes of the
specimens were 3 mm $\times$ 10 mm $\times$ 1 $\mu$m. A 20-T superconducting
magnet system (Oxford Inc.) was used for the dc magnetic fields, and a
two-channel nanovoltmeter (HP34420A) was used to measure the Hall
resistivity ($\rho_{xy}$) and the longitudinal resistivity ($\rho_{xx}$) by
using the standard dc five-probe method. The external magnetic field was
applied parallel to the $c$ axis of the thin films, and the transport
current density was $200\sim 250$ A/cm$^{2}$. Both the Hall resistivity and
the longitudinal resistivity showed Ohmic behavior, i. e., corresponding to
the flux-flow region, at the current used in this study.

\section{RESULTS and DISCUSSION}

We measured the longitudinal resistivities and the Hall resistivities of
Hg-1212 and Hg-1223 thin films in the magnetic field region 0 T $\leq H \leq$
18 T, and the results for Hg-1212 are shown in Fig. 1 while those for
Hg-1223 are shown in Fig. 2. Compared to most of previous experiments
performed at lower fields, we extended the magnetic field up to 18 T. The
motivation for doing this was to check whether the previous analysis of the
field dependence of the Hall conductivity based on the TDGL theory was valid
even at this high field. Figures 1(a) and 2(a) show the field dependences of 
$\rho_{xx}(H)$ for various temperatures. $\rho_{xx}(H,T)$ increases
monotonically with increasing temperature. In Figures 1(b) and 2(b), $%
\rho_{xy}(H)$ is plotted and has a nearly linear dependence on the field in
the high-field region. The sign of the Hall resistivity in the low-field
region near the transition temperature becomes negative, which is opposite
to the positive sign of the Hall resistivity for the normal state. The range
of the field in which sign reversal is observed for Hg-1223 is narrower than
that for Hg-1212. The insets of Figs. 1 and 2 show detailed representations
of the low-field region.

The Hall conductivity is typically defined as $\sigma_{xy}\simeq
\rho_{xy}/\rho_{xx}^{2}$ by assuming $\rho_{xx}\gg \rho_{xy}$. In Fig. 3,
the field dependences of the Hall conductivities of Hg-1212 (Fig. 3(a)) and
Hg-1223 (Fig. 3(b)) are shown for various temperatures. Based on the
theoretical prediction of Kopnin {\it et al.}\cite
{Kopnin1996,Kopnin1993,Kopnin1995}, we analyze $\sigma_{xy}(H)$ by using 
\begin{equation}
\sigma_{xy}(H)= \frac{C_{1}}{H}+C_{2}+C_{3}H ,  \label{sigma2}
\end{equation}
which is plotted with solid lines in Fig. 3. Compared to YBCO, the component 
$C_{2}$ is added for better fitting. The data are well fitted in the region
of 115 K $\leq T \leq 125$ K for Hg-1212 and 125 K $\leq T \leq 130$ K for
Hg-1223. In this figure, the downward curves, as approaching zero field,
show a sign reversal, but the upward curves do not. If the curve is
downward, $C_{1}/H$ is negative, but if the curve is upward, it is positive.

The temperature dependences of $C_{1}$ and $C_{2}$ for Hg-1212 and Hg-1223
are shown in Fig. 4. Experiment shows that $C_{1}$ scales with temperature
near $T_{c}$ as 
\begin{equation}
C_{1}\sim (1-t)^{n} ,  \label{c1}
\end{equation}
where $n$ is $2.3\pm 0.2$ for Hg-1212 and $1.8\pm 0.3$ for Hg-1223, as shown
in the Table I. The scaling form of $C_{1}$ can be partially understood from
the temperature dependences of the gap and of the coupling constant in Eq. (%
\ref{sigmaA}) based on the theoretical prediction by Kopnin. However, this
has not yet been proven rigorously. Scaling of $C_{1}$ has also been
reported for YBCO\cite{Ginsberg1995} and LSCO\cite{Matsuda1995} with $n$
values of 2 for YBCO and $2\sim 3$ for LSCO. These are not significantly
different from those for Hg-1212 and Hg-1223. Compared to several other
critical exponents, such as the magnetization scaling and the
irreversibility lines, $n$ does not critically depend on the anisotropy.

As shown in Fig. 4(b), $C_{2}$ steeply increases with decreasing
temperature. Therefore, we can extract the following scaling form for $C_{2}$
near $T_{c}$ : 
\begin{equation}
C_{2}\sim (1-t)^{n^{\prime}} ,  \label{c2}
\end{equation}
where $n^{\prime}$ is $3.2\pm 0.1$ for Hg-1212 and $2.0\pm 0.2$ for Hg-1223.
Differently from the case of YBCO, in which $C_{2}=0$, we find that $C_{2}$
is not negligible for Hg-based superconductors. $C_{2}$ seems to be
associated with the anisotropy ratio of the material, because the $C_{2}$
part of the resistivity becomes more explicit for the highly anisotropic
superconductors, such as LSCO\cite{Matsuda1995} and Tl-1212\cite
{Samoilov1994}. Specifically, a similar tendency to that shown in Fig. 4(b)
for $C_{2}$ was observed in data previously reported for LSCO, but the
temperature-scaling behavior was not determined. Not much information is
reported for Tl-1212; however, the Hall conductivity is well fitted by $%
\sigma_{xy}=C_{1}/H+C_{2}$. As explained before, the origins of the $1/H$
and $H$ dependences can be explained by the TDGL theory\cite
{Kopnin1993,Dorsey1992b} or the microscopic theory\cite
{Kopnin1996,Kopnin1995}, but neither of them can explain the scaling
behavior of $C_{2}$.

The coefficient $C_{3}$ of the term linear in $H$ shows a weak temperature
dependence in both Hg-1212 and Hg-1223. This is different from the cases of
underdoped and slightly overdoped LSCO\cite{Matsuda1995}, in which $C_{3}$
decreases as the temperature decreases. On the other hand, $C_{3}$ in YBCO
decreases linearly with temperature.

In order to investigate the effect of anisotropy on the coefficients $C_{1}$%
, $C_{2}$, and $C_{3}$ and on the powers $n$ and $n^{\prime}$, we summarize
our results along with previous results for HTS in Table I. As the
anisotropy increases, the absolute value of $C_{1}$ and $C_{3}$ decrease
while $C_{2}$ is increases. These values are evaluated at $t \simeq 0.92$.
According to this tendency, $C_{2}$ is very small for the case of low
anisotropy whereas $C_{3}$ is very small for the highly anisotropic case. As
a result, in Eq. (\ref{sigma2}), the second term is negligible for YBCO, and
third term is negligible for Tl-2212. In case of Hg-base superconductors,
however, since the anisotropy ratio ranges between that of YBCO and that of
Tl-2212, all terms in Eq. (\ref{sigma2}) are required, just as in the case
of LSCO\cite{Matsuda1995}. The Hall conductivities measured up to very high
magnetic fields (0 T $\leq H\leq 18$ T) for Hg-based superconductors are
still well described by Eq. (\ref{sigma2}), but the temperature dependences
of these coefficients have not yet been explained theoretically.

\section{SUMMARY}

We investigate the Hall effects for Hg-1212 and Hg-1223 thin films as
functions of the magnetic field up to 18 T. The Hall conductivity in the
mixed state is expressed well by $\sigma_{xy}(H) = C_{1}/H + C_{2} + C_{3}H$%
. The coefficient $C_{1}$ scales with temperature as $(1-t)^{n}$ with $%
n\simeq 2.3$ and 1.8 for Hg-1212 and Hg-1223, respectively; these values of $%
n$ are comparable to the values observed for YBCO and LSCO. We find that $%
C_{2}$ is more important for highly anisotropic compounds. $C_{2}$ is
observed to follow the same scaling form, but with exponent $n^{\prime}\sim
3.2$ and 2.0 for Hg-1212 and Hg-1223, respectively. These scaling behaviors
of $C_{1}$ and $C_{2}$ have not yet been explained theoretically.

\acknowledgments
This work is supported by the Ministry of Science and Technology of Korea
through the Creative Research Initiative Program.

% Fig. 1
\begin{figure}[tbhp]
\caption{The field dependences of (a) the longitudinal resistivity and (b)
the Hall resistivity for Hg-1212. The inset represents an enlargement of the
low-field region.}
\label{Fig1}
\end{figure}

\begin{figure}[tbhp]
\caption{The field dependences of (a) the longitudinal resistivity and (b)
the Hall resistivity for Hg-1223. The inset represents an enlargement of the
low-field region. The temperature and the field regions in which sign
reversal occurs for Hg-1223 are narrower than those in which it occurs for
Hg-1212}
\label{Fig2}
\end{figure}

\begin{figure}[tbhp]
\caption{The field dependences of the Hall conductivities of (a) Hg-1212 and
(b) Hg-1223 for various temperatures. The solid lines are fitting curves
obtained by using Eq. (\ref{sigma2}). For Hg-1212, the fitting ranges are
1.0 T $\leq H \leq 18$ T for 115, 120, and 125 K, and 2.2 T $\leq H \leq 18$
T for 100 K. For Hg-1223, the fitting ranges are 0.4 T $\leq H \leq 18$ T
for 125 and 130 K, 0.6 T $\leq H \leq 18$ T for 120 K, and 2.2 T $\leq H
\leq 18$ T for 110 K.}
\label{Fig3}
\end{figure}

\begin{figure}[tbhp]
\caption{The temperature dependences of the coefficients (a) $C_{1}$ and (b) 
$C_{2}$ in Eq. (\ref{sigma2}) for Hg-1212 and Hg-1223. The solid lines are
fitting curves obtained by using Eq. (\ref{c1}) for $C_{1}$ and Eq. (\ref{c2}%
) for $C_{2}$. The fitting ranges are $t=0.898\sim 0.984$ (114 K $\leq T \leq
$ 125 K) for Hg-1212 and $t=0.947\sim 0.985$ (125 K $\leq T \leq$ 130 K) for
Hg-1223. The fitting parameters are shown in Table I.}
\label{Fig4}
\end{figure}

\newpage \mediumtext
\begin{table}[h]
\caption{The data for $C_{1}$, $C_{2}$ and $C_{3}$ at $t\simeq 0.92$ for
various samples, where $\Gamma$ is the anisotropy ratio. For LSCO, the
values are for the optimal doping case with x=0.15 in La$_{2-x}$Sr$_x$CuO$_4$%
. The fitting formula $\protect\sigma_{xy} = C_{1}/H + C_{3}H$ is used for
YBCO, $\protect\sigma_{xy} = C_{1}/H + C_{2} + C_{3}H$ for LSCO, Hg-1212,
and Hg-1223, and $\protect\sigma_{xy} = C_{1}/H + C_{2}$ for Tl-2212. The
dash ($-$) symbols indicates 'not applicable' or 'not reported.'}\vspace{1mm}
\begin{tabular}{lcccccc}
& $\Gamma$ & $C_{1}$(T$/\Omega$cm) & $C_{2}$(1$/\Omega$cm) & $C_{3}$(1$%
/T\Omega$cm) & $n$ & $n^{\prime}$ \\ \hline
YBCO\cite{Ginsberg1995} & 50 & $-1.2\times 10^{4}$ & 0 & 250 & 2 & $-$ \\ 
LSCO\cite{Matsuda1995} & $1000\sim 1500$ & $-$ & $\sim 14$ & $\sim 80$ & $%
2\sim 3$ & $-$ \\ 
Hg-1212 & $-$ & $-540$ & 160 & 31 & $2.3\pm 0.2$ & $3.2\pm 0.1$ \\ 
Hg-1223 & 2500\cite{YCKim1995} & $-300$ & 210 & 55 & $1.8\pm 0.3$ & $2.0\pm
0.2$ \\ 
Tl-2212\cite{Samoilov1994} & $10^{4}$\cite{clinton1995} & $-119$ & $-$ & 0 & 
$-$ & $-$%
\end{tabular}
\end{table}

\end{document}